# Integrated Agent-based Microsimulation Framework for Examining Impacts of Mobility-oriented Policies

Muhammad Adnan[a*], Fatma Outay[b], Shiraz Ahmed[a], Erika Brattich[c], Silvana di Sabatino[c], Davy Janssens[a]

[a] *UHasselt - Hasselt University, Transportation Research Institute (IMOB), Agoralaan, 3590 Diepenbeek, Belgium*
[b] *Zayed University, Dubai, 19282, UAE*
[c] *Department of Physics and Astronomy, Alma Mater Studiorum, University of Bologna, Bologna, Italy*

**Abstract**

Travel demand management measures/policies are important to sustain positive changes among individuals' travel behaviour. An integrated agent-based microsimulation platform provides a rich framework for examining such interventions to assess their impacts using indicators about demand as well as supply side. This paper presents an approach, where individual schedules, derived from a lighter version of an activity-based model, are fed into a MATSIM simulation framework. Simulations are performed for two European cities i.e. Hasselt (Belgium), Bologna (Italy). After calibrating the modelling framework against aggregate traffic counts for the base case, the impacts of a few traffic management policies (restricting car access, increase in bus frequency) are examined. The results indicate that restricting car access is more effective in terms of reducing traffic from the network and also shifting car drivers/passengers to other modes of travel. The enhancement of bus infrastructure in relation to increase in frequency caused shifting of bicyclist towards public transport, which is an undesirable result of the policy if the objective is to improve sustainability and environment. In future research, the framework will be enhanced to integrate emission and air dispersion models to ascertain effects on air quality as a result of such interventions.

## 1. Introduction

Collective human travel behavior is strongly linked with deteriorating air quality, global warming, environmental noise and many other issues concerning quality of life. The transport sector has contributed around 23% of the total $CO_2$ emissions, which is the 2$^{nd}$ highest in Europe (EEA 2016). Previous studies indicate that a variety of transport demand management strategies can change/manage activity-travel behavior of individuals to promote sustainable transport options, which are vital for air quality improvement (De Nazelle et al 2011). In many European cities, transport demand measures (such as congestion charging, fare reduction, increased parking prices, low emission zones, etc.) are already rending fruitful results. The first order impact of the above-mentioned interventions/policies on the individual activity-travel behaviour can be in the form of change in the choices of mode/ route/ departure times or even change in the sequence of out-of-home activities. Because of the changed behaviour of the individuals in an area, the second order impact of these interventions could be an improvement in the overall air quality of the city (Ben-Akiva and Bowman 1998, Adnan et al 2010, Auld and Muhammadian 2012). This improvement in air quality can reduce the exposure of citizens. It is also necessary to further refine these interventions or come up with more innovative interventions that provide optimal results. Implementing such interventions is difficult and therefore, the use of a simulation platform provides a flexible tool to ascertain impacts of interventions.

Activity-based microsimulation models (ABM) are more detailed and keep track of individuals considering all aspects of an individual activity-travel [*Rasouli and Timmermans, 2014*]. A typical ABM considers the complete daily activity-travel pattern of individuals living in the study area. This includes for each agent in the synthetic population, the number of activities to be performed and specific attributes of each activity: type, start time, duration, and location. Furthermore, these simulated activities are also linked together via a travel component having its dimensions: travel time and travel mode. However, these ABMs can only predict activity-travel schedules and require assignment model for defining routes of travel. Integrating the activity-based model output with traditional traffic assignment model may cause loss of entire detailed information along with individual identity on the road network [*Lu et al., 2015*]. An appropriate integration can allow this information to be intact when an individual activity-travel pattern is executed on the network. MATSIM [*Horni et al., 2016*] and SimMobility [*Adnan et al., 2016*] are two examples of open source platforms that provide such a framework. The documentation available for MATSIM is more detailed and it is being used quite widely by the transportation researchers and planners (Kay et al 2016). This paper presents an application of the integrated simulation platform that uses a relatively lighter version of activity-based model (ABM) and MATSIM model to simulate mobility-oriented policies for two cities in Europe, namely Hasselt in Belgium and Bologna in Italy. The focus of the paper is to discuss the impacts of two policies (i.e. restriction of car access on some specified network links and increase in bus frequencies on particular routes); however, a few details of the integrated model and its calibration is also presented to better understand the results. The lighter version of the ABM is based on the notion that it provides all relevant information required by MATSIM, even though, on the other side, it contains two behavioral models; 1) mode choice for work and school activities and 2) a joint mode-destination choice model for shopping and leisure activities.

---

[*] Corresponding Author, Tel.: +32-11-269-147 ; Fax: +32-11-269-199 ; *E-mail address*: muhammad.adnan@uhasselt.be

The rest of the paper is organized as follows. In Section 2, we give a brief overview of the related existing work especially from studies that examined the impacts of transport demand management (TDM) measures. In Section 3, we discuss the simulation framework. Section 4 provides the results of the policies implemented in the two previously mentioned European cities. Section 5 discusses the results followed by a conclusion section of the paper.

## 2. Related Existing Work

Previous studies have been reporting a continuous growth in the ownership of motorized vehicles . This has caused a significant threat to healthy life style as related to a worsening of air pollution and climate change [*Ahmed et al 2019*]. Human exposure to air pollutants is largely due to the traffic and transportation activity, which emerges as a direct link between respiratory problems (especially residents living around busy roads), increased allergies, birth defects and numerous forms of cancer [*Int Panis et al., 2016*]. Despite significant efforts, a significant number of city residents in Europe are still exposed to levels of air pollutants exceeding EU air quality standards [*EEA, 2016*]. It has been mentioned that implementation of EU Clean Air Policy framework by 2030 can evade more than 58000 deaths along with this it also brings savings of around 40 billion Euros in the health sector. Vehicle emissions from the transportation activity also contribute to changing climate, with a share of greenhouse gas emission from this sector estimated to be around 23% (EEA 2016). It is further stated that this contribution is growing at a much faster rate. EU legislation '20-20-20 by 2020' [*Landis et al., 2012*] proposed a reduction of greenhouse gas emissions of at least 20% by 2020 compared to levels of 1990 across all sectors. In this regard, reduction in on-road emission through change in travel behaviour has been suggested as the most effective way for achieving it [*De Nazelle et al., 2011*]. Transport demand management is also regarded as an attractive strategy for reducing environmental impacts because of its inherent nature in addressing multiple impacts simultaneously, as demand for transport is derived from a large range of sources.

### 2.1. Public Transport related interventions and their impacts

The European Commission's Green Paper on Urban Transport [*UTIP, 2008*] has realised this fact long ago, and therefore advocating policies that encourage higher use of public transport. Benefits in investing and promoting public transportation system to reduce car travel are no doubt beyond comprehension. Many studies have attempted to quantify these benefits in terms of economical (i.e. reduced vehicle miles) and environmental (improved air quality) cost savings, and further translated these benefits into health impacts. Previous studies further concluded that to reach a long-lasting effect of the public transport system, it should be embedded in an integrated approach of multi-modality and cooperation with cycling, car share and other mobility services (*Fellermann* [*2015*]). *Cairns et al.* [*2004*] reported that improvement in public transport services alone is not able to attract the desired number of clients, however, with an efficient informational campaign the number of bus journeys increased by around 42%. *Koo et al.* [*2013*] examined the effects of point card system employed in South Korea to promote green growth (i.e. more use of public transport) using stated preference data. Their results indicate that consumer low membership cost and use of accumulated points are the key factor for widespread participation. Using a choice behavioural model estimated for Portland, Oregon metropolitan area in USA, *Shiftan and Suhrbier* [*2002*] investigated the changes in travel behaviour by improvement in transit service (through increase supply of buses and reduction in fare). The results of their model suggest an overall increase of 10% in bus ridership, even though only part of this increase derived from people who previously used car while the rest originated from other high-occupancy or non-motorized modes (cycling and walking).

*2.2. Car restriction and road pricing related interventions and their impacts*

Low emission zone (LEZ) is a monitored and defined area where access restrictions are employed for certain vehicles that do not meet emission class criteria, but where restricted vehicles can enter the area by paying a heavy monetary penalty (In case of London, the penalty is around 100-200 GBP per day for large vehicles) [TfL, 2008]. The enforcement of the LEZ is usually done through an Automatic Number Plate Recognition (ANPR)-camera system. In Europe LEZ is a common intervention (classified as structural strategy) employed to reduce vehicle emissions to improve air quality [Malina and Scheffler, 2015]. However, each LEZ in the different regions of the EU has its specifications in terms of restrictions, monitoring and enforcement mechanism and penalties. There is a variety of added advantages reported for LEZ along with improvement of air quality, such as:
- Reduction in congestion within the LEZ area (usually these areas are internal areas of the cities where congestion is itself a significant problem), though in the longer run other methods are required to maintain traffic volume levels.
- Gradual replacement of polluting vehicles into low emission vehicles (especially positive changes in the composition of fleet vehicles).
- More availability of parking spaces inside the LEZ, as restricted vehicles will tend to use Park & Ride option outside the LEZ area to avoid heavy penalties.

The major behavioural response about LEZ was noted in terms of change in ownership of vehicles allowed to enter the restricted area. The majority of these ownership changes are noted in commercial/freight vehicles [*Ellison et al., 2013*]. None of the previous studies estimated the secondary effect of this intervention i.e. changes in travel behaviour (in terms of changing location choice for conducting various activities, use of different transport modes to enter the LEZ). However, this intervention was found to produce a significant effect in terms of improving air quality (as depicted from REFF). Because of these significant positive impacts, recently the implementation of this intervention was planned in various other European cities such as Antwerp (Belgium) and Prague (Czech Republic) [*EU, 2016*].

Road pricing can be regarded as another variant in relation to access restriction. It is being employed in many regions of the world with a view point that road users should pay for using the road network for correct allocative decisions between transport and other activities. *Ubbels and de Jong* [*2009*] in their comprehensive review of pricing strategies mentioned that road users respond in various ways in terms of their behaviour against road pricing which could be changed in all or some activity scheduling dimensions such as route, time-of-day, location, mode or changes in activity patterns along with long terms choices such as vehicle ownership, residential and work location, etc. Further, it is mentioned in the literature that different types of pricing methods can have different impacts, like for example: tolls may affect travel routes and destinations, a time-based pricing may shift trips to other times. The increase in fuel prices tend to affect the type of vehicles purchased more than vehicle mileage. In addition, the type of trip and traveller is also important, for example: commute trips are found to be less elastic compared to recreational or shopping trips; individuals with high income are less sensitive compared to low income individuals. In Singapore, implementation of time-based pricing system in the central business district (CBD) resulted in: (1) reduction of 73% of private vehicles entering CBD, (2) increase in carpooling by 30%, (3) doubling of the bus usage. However, these results are achieved with the implementation of other measures such as strict conditions on vehicle ownership and improvement of public transport operations [*Schuitema, 2007*]. Studies regarding the implementation of congestion charging schemes reported similar findings. In London (UK), a time-based area wide scheme resulted in a 30% congestion drop after 1 year, even though the reduction in delays was just around 8% after 3 years [*TfL, 2007a, TfL, 2007b*]. In Stockholm (Sweden), a similar scheme resulted in a 20% reduction of traffic (in 2011) compared to based year (2006) [*Ubbels and de Jong, 2009*]. Simulation models were also used to test a wide range of scenarios related to road pricing. Wu et al (2016) investigated the impacts of various methods of congestion charging i.e. distance based or cordon based, for different sizes of area in Beijing (China). In another study conducted by [*Guzman et al., 2016*], the testing of a combination of strategies including road pricing, increased parking fees and fuel taxes using a land use and transport interactive model for Madrid Region indicated that pricing measures in a local or central area are helpful to achieve local target of air quality, even though they can cause negative impacts over surrounding areas. The use of Integrated simulation tools for intervention assessments are very limited owing to associated complexities.

This study aims to examine an application of an already developed integrated simulation platform that uses a lighter version of the activity-based model and integrate it with MATSIM. Base cases for two European cities are calibrated and interventions for car access restriction and improvement of public transport infrastructure are examined and compared with other similar studies.

## 3. Integrated Microsimulation Framework

### 3.1. Integration Mechanism

The integrated microsimulation framework is based on an Activity-based approach and MATSIM simulation platform [*Horni et al., 2016, Ziemke et al., 2015*]. The model cyclic chain starts from the synthetic population (containing socio-economic properties of person and household) of the city, predicting their travel routine (in the form of activity-travel diary) and then executing it on the transportation supply network to obtain traffic volume on the links for an entire day. This is done in a way that individual properties are kept intact at the supply side. The MATSIM structure is such that it requires population schedules (also known as plans) that are executed in a simulation platform. MATSIM follows an iterative algorithm, where not only route assignment process is employed but also scores of alternative plans of an individual are calculated (the alternative plans are generated by changing activity start times, activity duration, modes for trip, activity sequence and routes based on an input plan of the individual). MATSIM employs a co-evolutionary algorithm to optimise the plan based on the network situation and attempt to establish an equilibrium. Usually, 250 or more simulations runs are recommended to achieve optimized results. It should be noted that MATSIM does not provide any visualization capabilities to examine the results and requires further processing of the results to get the desired outputs. In addition, usually a portion of the population (selected randomly, around 10% to 30% depending on population size) is simulated within MATSIM to allow for the lower run time. The results are then extrapolated further. Based on MATSIM output, new skim matrices are generated (travel times on zonal basis) and then are used in the lighter version of ABM to produce the schedules. This cycle is then run a few times to reach a global equilibrium. Figure 1 presents the integrated microsimulation framework to simulate the two previously mentioned cities. Some important datasets that are used in the model are as follows:

- OpenStreetMap (OSM) data for the two (Bologna and Hasselt) cities. This data is then converted after necessary cleaning to provide a road network to MATSIM.
- General Transit Feed Specification (GTFS) data for all cities are available, containing information about public transport routes, stops and their schedule in a standard format. This data gives the necessary details to establish public transport network as an input to MATSIM after employing the necessary cleaning and integration process with OSM data [*Vurrstaek et al 2018*].
- Traffic volume data on important links of two cities is used for calibration of the final output from MATSIM.

The calibration process is developed in a way to change only a few important parameters within the MATSIM configuration file and road network parameters. The parameters used for optimization are related to mode utility score function (mode params) with the aim that travel mode use distribution in the study area is satisfied. Along with these parameters, other sets of parameters used for calibration are link capacities (provided in the road network data) and parameters related to early departure and late arrival. In addition, a few parameters are used from the two models that are part of the light version of ABM. The objective function is the ordinary least square (OLS) type that minimizes the difference between the observed traffic volume and predicted traffic volumes (only cars) on links where data is available. The algorithm used is SPSA (simultaneous perturbation stochastic algorithm) to find out the new set of parameters to be used for the next calibration iteration as used in [*Oh et al 2019*]. It should be noted that within one calibration simulation there are 250 MATSIM simulation runs, and therefore, the process is quite time-consuming.

The Iterative Proportional Updating (IPU) algorithm has been used to generate a synthetic population. The algorithm code has been developed at IMOB, Uhasselt, in the Python language. This algorithm was first developed by [Ye et al., 2009]. The algorithm is for matching both household and person marginals by updating sample household weights. In the fitting step, household and person type constraints are estimated using the fitting procedure, followed by the calculation of sample household weights by the IPU algorithm [*Cho et al., 2014*]. Based on the household selection probabilities, households are sampled and expanded and finally, the population of households is generated. Sampled household weights generated in the fitting steps are used to generate household selection probabilities. The algorithm requires two types of data: seed data and target marginal information. The seed data contains disaggregate population which normally describes enough details of the population (e.g. household and personal information and their considered socio-economic characteristics), but with only a small number of individual elements in the seed data. On the other hand, target marginal only has information about the sum of attributes in a spatial unit e.g. total number of males, total number of females, number of individuals in a particular age category, number of households with one car etc. The goal of IPU is to use the seed data in a way that it expands the sampled household and persons to fit into all considered target marginal attributes of a particular TAZ.

*3.2. Lighter Version of ABM*

In absence of detailed activity-travel diary data for cities, we have developed a light activity-based model based on certain rules and assumptions for estimating schedules for the population. The model segments of the population are based on a few characteristics and compared with the obtained schedules of similar population segments of the Flanders region. After that the model draws an activity sequence from these schedules. These activity sequences are further enriched with other scheduling dimensions (such as activity duration, modes, departure times etc.) based on the activity-travel diary data recorded from two cities and some published travel research in those cities. Some consistency checks are performed to see the travel mode shares, trip distance and time-of-day distributions are consistent with existing trends in the available aggregate data. The output of the activity-based model is then used as input to MATSIM by incorporating another procedure to obtain node-based activity locations from zonal-based locations. The node is part of the network used to perform MATSIM simulation. More details of lighter version of ABM are as follows:

- Activity schedules are obtained from an already existing full-scale operational activity-based model for the Flanders region in Belgium. These schedules (in terms of activity sequence, such as home-work-home or home-work-shopping-home) frequencies are grouped into the type of population characteristics of Flanders (Gender, Age, Family size, Income, etc.) that are available for two European cities (Hasselt and Bologna), developed as multivariate joint distribution. For example, for a specific value of gender, age and family size, relative frequencies are estimated for various types of activity sequence patterns.
- Using a Monte Carlo simulation method, for Bologna, a synthetic individual having a particular population characteristic has been assigned to a particular activity sequence type.
- Once a sequence type is assigned, for each activity type distribution of duration, activity start time is derived from the available activity-diary (GPS based smartphone data) of a few individuals from Bologna and Hasselt. Based on these distributions, again using a Monte Carlo simulation method, duration and activity start time are allocated to each activity in the activity sequence.
- A tour-based mode choice model for work and education activity types are developed from the smartphone activity travel diary data for each city. Skim matrices information is derived from Google Maps API and available datasets for the region. The model followed a multinomial logit model structure and estimated using Biogeme®. From the activity-diary data set, tours for different activities are formulated following a methodology used in [*Lu et al 2015; Siyu 2015*]. Using this model, travel mode distribution for different activities in a tour is predicted. A travel mode is then assigned for a trip between two activities of a schedule.
- Similar to the tour-based mode choice model, joint mode-location choice tour-based models are developed for shopping and other activities utilising the smartphone data, available land use information, skim matrices and other personal characteristics. The tour-based mode choice model and joint mode-location choice models are the two main components incorporated into the integrated framework to represent behavioural notions (in full-scale ABM model there a variety of other models exists such as time-of-day models, intermediate stop mode, location and time-of-day models, activity pattern model, etc).

- The next major step is to allocate activity location zones. Based on the residential point of interest (POI) density of each zone, all individuals in the synthetic population are assigned a particular zone (for home location) in the respective city [*Baqueri et al 2019a, 2019b, 2018*]. A similar approach is followed for allocation of work activity location in terms of zones. For secondary activities of the tour (e.g. shopping, other(leisure, recreational)), again a similar method is followed to allocate zone from the set of candidate zones, The candidate zones, however, are ensuring that the zone location is within the threshold limit (i.e. 10 km for Bologna) from the origin zone. This is to avoid giving farthest zone to the secondary activity of the tour.
- Some consistency checks are carried out to finalize the schedule, as described in the following:
  - Bicycle and walk trips are not more than 8km and 2 km respectively.
  - If the personal vehicle (i.e. car/bicycle) is selected for a particular tour, it is ensured that all other trips of the tour are carried out with the same mode.
  - It is also ensured that cumulative mode shares for these schedules followed other sources of data available from [*Comune_di_Bologna, 2019*].
  - Activity duration of the second last activity in the pattern is adjusted such as the last activity *home* can be started as late as 3:00 am next day.

## 4. Application of integrated microsimulation framework

These studies were conducted as part of the iSCAPE H2020 project, where different project activities aiming for to find solutions for improving air quality and mitigating climate change are carried out in six European cities i.e. Bologna (Italy), Bottrop (Germany), Dublin (Ireland), Hasselt (Belgium), Guildford (UK) and Vantaa (Finland). Therefore, with the collaboration of two partner's organizations (UH and UNIBO), and in contact with the respective Municipalities and Environmental Protection Agencies in Hasselt (Belgium) and Bologna (Italy), the required datasets previously described were gathered. The results obtained are presented and discussed in the following sections.

*4.1. Hasselt City*

Hasselt city is located at the junction of important traffic arteries from several directions. The most important motorways are the European route E313 (Antwerp-Liège) and the European route E314 (Brussels-Aachen). The old town of Hasselt is enclosed by two ring roads. The outer ring road serves to keep traffic out of the city centre and the main residential areas. The inner ring road, the "Green Boulevard", serves to keep traffic out of the commercial centre, which is almost entirely a pedestrian area. There are also important traffic arteries to Tongeren, Sint-Truiden, Genk, and Diest. In Hasselt, 73% of the total trips are performed using a car (Stad_Hasselt 2019). For a trip distance of around 8 to 10 km, the bicycle has the 2$^{nd}$ largest share after the car, which is the result of adopting a bicycle oriented policies by the city administration such as the development of relevant infrastructure (i.e. segregated bicycle lanes, bicycle priority streets, bicycle parking facilities), the availability of bike sharing schemes and the increase in parking cost in the inner core areas [*Stad_Hasselt, 2019*]. However, in the inner core region, there are streets which still allow car movements despite some of the shopping streets which are completely pedestrianized. Public transport is available, and a major transport hub is located near the Hasselt train station. There is only one operator (*Delijn*) providing the bus-based public transport services within the Hasselt Arrondissement. However, bus service is not very frequent except for a few routes that cover Hasselt train station and Hasselt University campus located in Diepenbeek (a small town near Hasselt city).

The MATSIM simulations were performed by selecting a geographical region including Hasselt city. In this manner, from the entire Flanders population only those individuals with at least one activity located within the geographical region as available from Flanders schedules are considered. This means that the simulation process includes all Hasselt population along with those individuals who are not residing in Hasselt but whose schedules predict an activity located within that region. Through this option, all external trips within Hasselt region are also captured, a feature which is of prime interest. Each MATSIM simulation run (that include 250 iterations) is performed in 612 minutes. The calibration took slightly longer than 20 days of a server machine (3.50 GHz with 12 cores) to run those 50 iterations. The staring value of RMSE is 1.8 and after 50 iterations it has been lowered to 0.15, which can be considered sufficient for such a complex calibration framework [*Lu et al., 2015*]. Further lower values are desirable, but given the time limits, the calibration simulation was restricted to 50 as there seems to be very little improvement when proceeding to longer iterations.

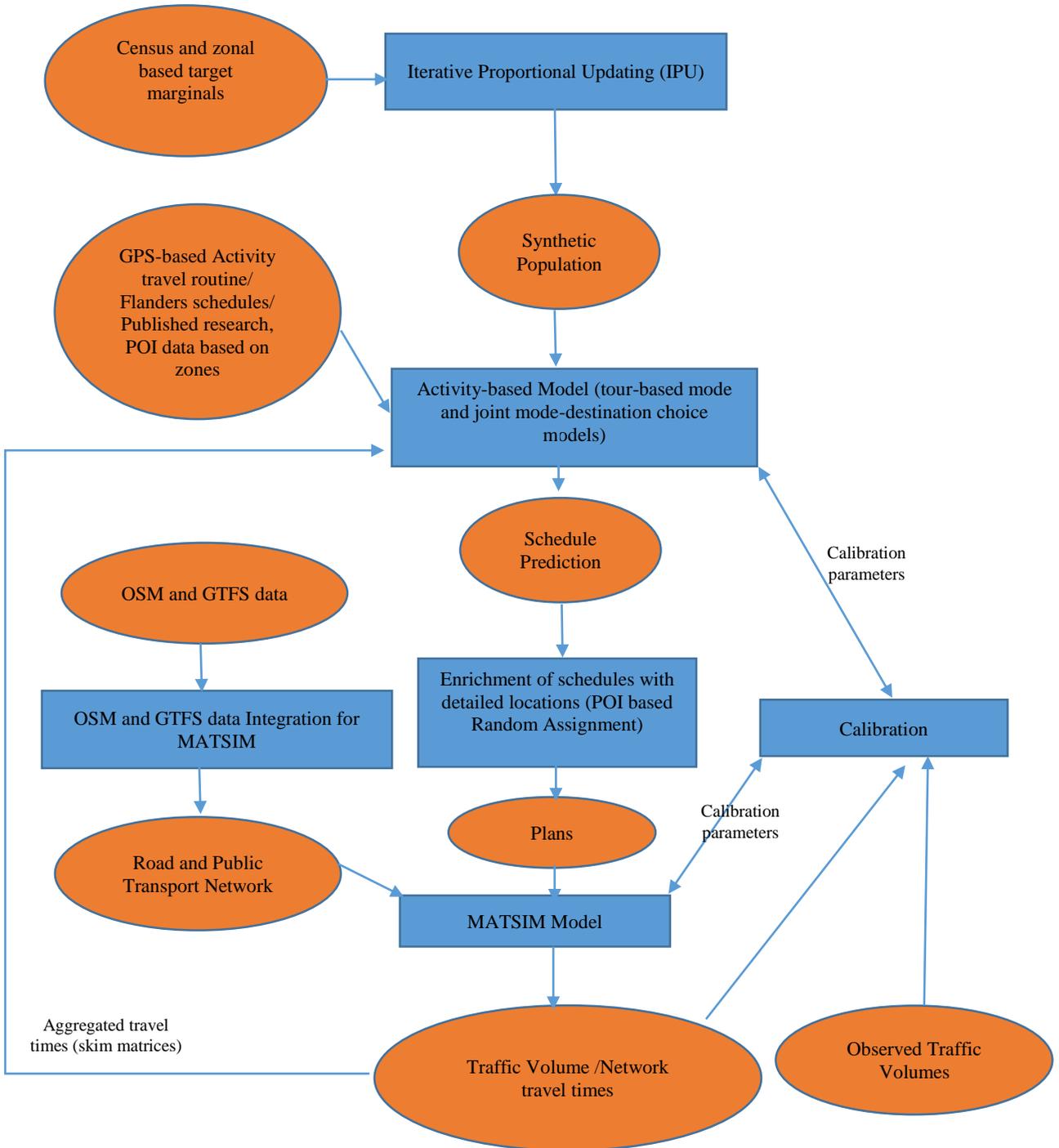

Figure 1: Integrated microsimulation framework

The traffic volume (including freight traffic and public transport) for the morning peak hour (base case) can be observed in Figure 2a for the reduced road network (representing only Hasselt city premises). Most of the roads contain a traffic volume between 150 to 1500 vehicle/hr (which is lower than the capacity limits), or less. However, there are several links which are in yellow and orange colour, clearly showing that traffic volume in the outer ring in rush hour is higher than the capacity limits. The traffic volume is large especially at the main junctions of the outer ring and at the point where the outer ring connects to the E313 motorway ramps (in the southern part of the outer ring) linking Hasselt to the other major cities of Belgium (such as Brussels, Antwerp, Liege etc.). It should be noted that in the inner ring of Hasselt, there is also significant traffic, even though this is considered as a green boulevard. Furthermore, all the major radial roads connecting the inner ring also present significant traffic volume. Most of these major radial roads are connected to links where there are schools. Furthermore, these connecting roads are attracting traffic from the residential roads in between the inner and the outer ring, presenting less traffic as compared to the major radial roads that provide smooth connection between the inner and the outer ring.

### 4.1.1 Car Access Restriction

The policy is mostly implemented in many cities of the world by restricting access of diesel cars, or allowing access by paying certain fees (congestion charge zone). To raise awareness of citizens, Hasselt city organised on an annual basis an event in the city where for a single day the inner ring and all roads around it are completely banned for cars. Keeping this as the context, we tested an intervention where 319 links in the road network are completely restricted to access from cars. However, the buses and taxis are allowed to access in the zone. The policy is implemented by incorporating changes in the config.xml file of MATSIM, where another component usually used to test road pricing policies [*Horni et al., 2016*] is added. In order to restrict completely the car access to those links identified in the road network, we inserted a very high toll to travel on these links. As such, the utility of choosing a route that contains any of those links has a very low value as compared to the other routes in the network, or it may also cause change in travel mode or some other scheduling dimension. Figure 2b presents the traffic volume for the morning peak hour for this case to be compared to that of the base case (i.e. Figure 2a).

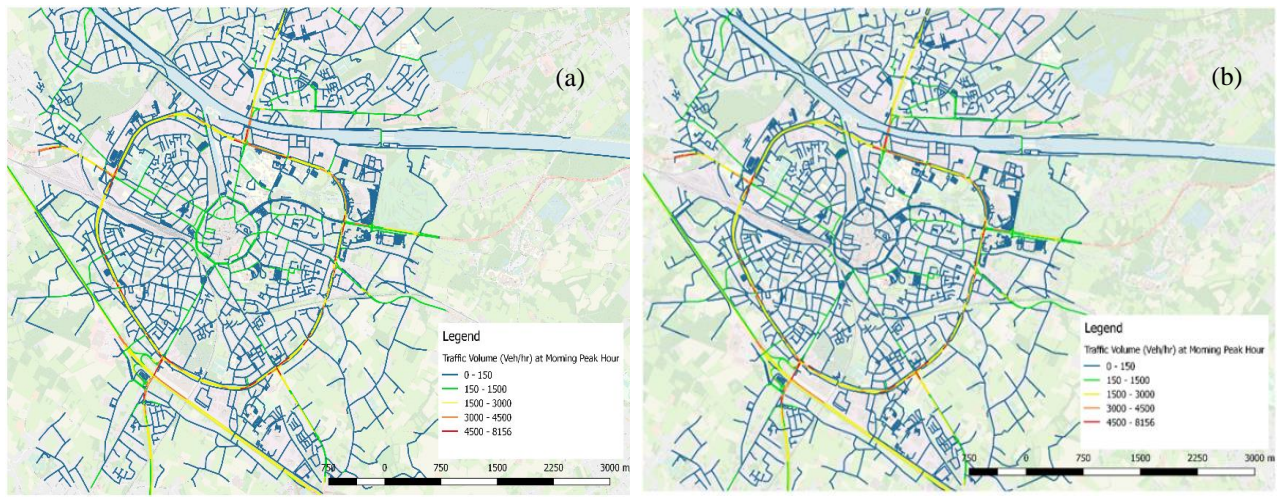

Figure 2: Morning peak hour traffic volume (vehicles/hr) on reduced Hasselt Network- (a)Base case (b) Car Access Restricted Policy

It can be observed from Figure 2(b) that the inner ring is now in the lowest category of traffic volume (the traffic volume is only due to the buses on those restricted links). Other major roads that connect traffic from the inner to the outer ring are in the 2$^{nd}$ lowest category (where traffic volume is still under capacity limits). However, the close examination of values of traffic volume on these radial roads revealed that traffic volume is much less on these roads compared to the base case. This is because only the local residents are now using these radial connectors to go towards the outer ring to find the best possible route to their destination. Furthermore, there is no notable difference found on motorways and other primary arterial roads. In addition to this, we observed the following major changes:

a) Several drivers have detoured their routes as they are now using longer routes in comparison with the base case where car drivers were using more direct routes. It is noted that the travel time of the car drivers has been increased by 9% on average
b) Due to the introduction of a very high penalty for using the inner ring, 22% of the car trips where the destination was the inner core zone of the city were reduced. This has been distributed as 11% to public transport, 7% to bicycle and 4% to walking. Furthermore, around 6 - 7% of individuals have changed their location of activities from the inner city areas to other areas in Hasselt city.
c) The majority of the individuals that have changed their travel mode because of this policy belong to the retired and student population.

Figure 3 presents the overall mode share differences between the base case and this policy. The largest difference in percentage points (2.1%) are noted for car passenger use followed by car drivers, and it is translated in terms of use of more buses and then bicycles. In relation to absolute numbers, PT trips increased from 10,150 to 14,385 trips because of this policy (i.e. a difference of 4,235 trips in a day) and car_driver and car_passenger trips decreased from 138,200 to 131,370 daily trips ( i.e. a decrease of 6,830 car trips/day). This is a reasonable number of decreases in car traffic compared to the size of Hasselt city, and may be helpful in improving the air quality.

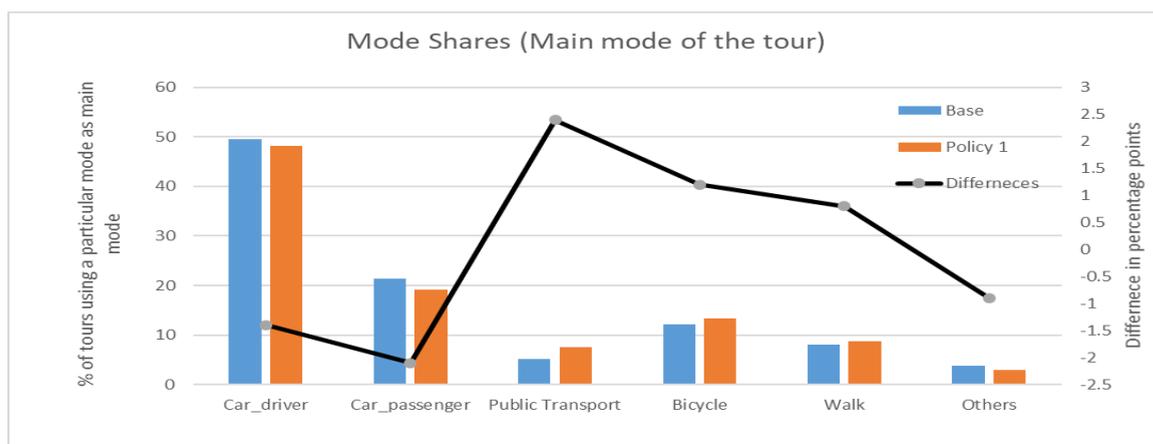

Figure 3: Mode shares as main travel mode of the tour (along with differences with base case)

### 4.1.2 Enhancement of Bus Services

Hasselt city concentrates relatively less investments and subsidies for bus-based public transport in the city. As a result, operators have reduced their bus fleet to a significant extent and recently most of the bus routes were re-designed. More than 80% of the households have car ownership and public transport (PT) is mainly used by low-income individuals, students and senior citizens. All these categories of citizens also have to pay less fare in comparison with other classes. Additionally, children under 12 years of age are enjoying a free ride on the bus. To encourage the use of public transport, Hasselt city council is now extending this facility to individuals up to 20 years [*Stad_Hasselt, 2019*]. However, in the current situation, only 6% of tours have their main mode of travel based on PT, and one of the prime reasons is the reduced frequency (higher waiting times) of the buses. In total, within Hasselt city, there are around 50 bus routes including 22 dedicated for serving areas within Hasselt and other bus routes that are serving part of the area of Hasselt and continues towards other surrounding cities (such as Genk, Tongeren, Sint-Truiden, Masseik and even Maastricht (a city in Netherlands) etc.). In peak hours during regular weekdays (also school days), only a handful of routes have a frequency of 5 minutes (routes serving Hasselt university campus at Diepenbeek), and some have frequency of an hour as well. However, during the non-peak hours the frequency of the buses is low (around an hour), and therefore individuals tend towards other modes of travel. There is a single operator Delijn that operate the buses. There is also a smartphone application from Delijn, which not only provides the schedule of bus services but also a functionality to pay as you go, in case individuals do not have their regular bus passes. A single journey fare is a bit expensive if paid by cash as compared to electronic payment via smartphone, the fare difference being almost 40% [*Dlijn, 2019*]. In addition to this, there are monthly, 3 monthly and yearly passes available for low income individuals, where basic health insurance companies cover the expenses of the bus travel. Apart from this, several employers within the region provide bus fare subsidies to their employees. Public transport users mostly perceived the bus service as less reliable especially in non-peak hours and sometimes in peak-hour as well. Furthermore, it has been also realised that the bus routes have very little coverage and at the same time a quite low frequency in places where most of the University students are living. Based on this context, in the microsimulation platform a scenario of increase in the frequency of buses was implemented. We increased the frequency of around 50% of the current bus routes by double and by 25% for other 50% bus routes. Those bus routes where the bus service is already 5-10 minutes frequent in peak hourswere not enhanced. Based on this, we have estimated that the bus fleet size requirement will be increased by 35% and at the same time *Delijn* may need to employ more bus drivers and other ground staff, which is a significant investment. The most crucial component to run this policy is the mode choice model within ABM and MATSIM. Within the mode choice model, the PT utility contains a waiting time variable, whose values are changed during the execution of this policy scenario.

The results indicate that this change increased the PT mode share by only 4.8% points (i.e. the overall share of PT for this scenario is 10%). Because of this scenario, car use has been reduced but it is not significant as the majority of the bicycle and on-foot travellers have shifted their mode to PT rather than car users. Figure 4 provides a clearer summary of the changes in mode share distribution in this situation. In absolute terms, PT trips increased from 10,150 to 19,520 (i.e. an increase of 9,370 trips in a day), even though this increase is at the cost of a decrease in bicycle trips from 23,815 to 20,496 and walking trips from 15,616 to 13,100. Car traffic is reduced by only 1.8% points, which is around 3,000 trips in magnitude. The major reason could be that even with low waiting times, the bus is still considered as a not attractive travel alternative for individuals who are captive car users.

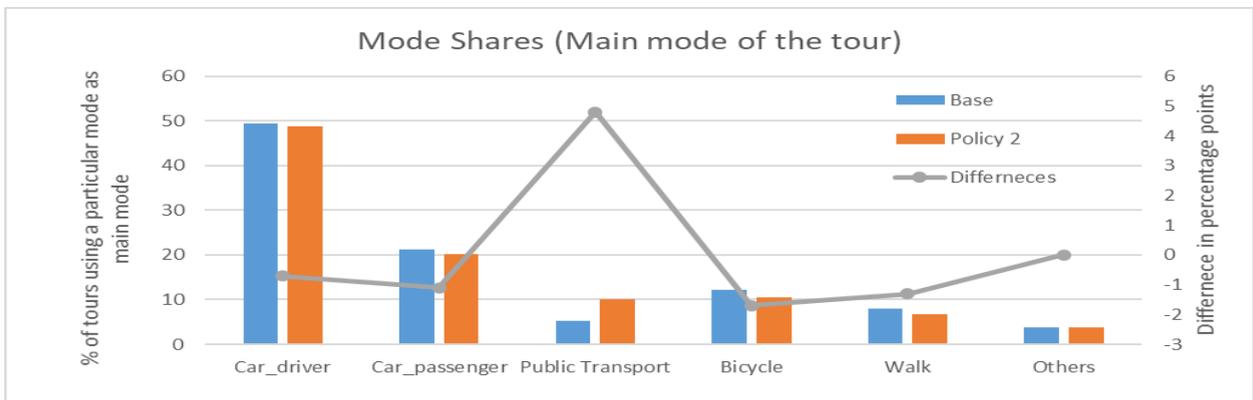

Figure 4: Mode shares as main travel mode of the tour (along with differences with base case)

*4.2. Bologna City*

Bologna is the seventh largest densely populated city in Italy. It is located in the Po Valley in northern Italy at the foot of the Apennines and between the rivers Reno and Savena. It is the capital and largest city of the Emilia-Romagna Region with a population of 388,567 inhabitants. It has a population density of 2766 per km$^2$ [*CityPopulation, 2019*]. Its location at the crossroad between north and south of Italy, and the presence of the international airport makes the city as a major transportation hub. Bologna is also known as a University town as it is one of the oldest universities in the Western world with around 85,500 students [*de Ridder-Symoens and Rüegg, 2003*]. Due to the city's strategic location as a crossroad between north-south and east-west routes, Bologna central train station is also very busy, serving 58 million passengers annually (Tositti et al 2014). By road the city is accessible via the A1 from Milan, the A22 and E35 from Verona and the A13 from Venice. Due to the industrial and educational zone, lot of commuters enter/ leave the city and create traffic congestion especially on the approach road to the cities on peak hours. Although the city is served by a large network of public bus lines, including trolleybus lines, operated since 2012 by Trasporto Passeggeri Emilia-Romagna SpA (TPER), the highest mode share for transport in the city is by car i.e. 35%, followed by public transport i.e. 26% [*Barrett, 2018*]. The average amount of time people spends commuting with public transit, for example to and from work, on a weekday is 53 min. 9% of public transit riders ride for more than 2 hours every day. Keeping in view the increasing traffic volume on roads approaching toward the city centre, the city of Bologna is taking measures and designing policies to tackle this problem. More than 10 parking areas below are created for the commuters who come via car to visit the city centre and surrounding area for performing various activities. The inner core of the city is also made traffic free and the area is open only by foot or bike (limited traffic zones). City authorities are further interested to test scenarios for extending the restriction of car access and facilities opening times.

For Bologna, the links counts are available for another calibrated network that includes only major roads. Furthermore, the link counts available for the Bologna network contains total link volume (combining both directional flows). Therefore, for calibration purposes, only important links that are part of the two networks and total volume (in both directions) are used. The final output is expanded for the entire population. Each MATSIM simulation run (that includes 250 iterations) is performed in 432 minutes. The calibration took little longer than 15 days of a server machine (3.50 GHz with 12 cores) to run those 50 iterations. The staring value of RMSE is 2.4 and after 50 iterations it has been lowered to 0.26, a value which can be considered satisfactory.

4.2.1   Car Access Restriction

In this policy, we have extended the network of car free links, to see what improvements can be achieved in terms of air quality. Keeping this as the context, we tested an intervention where 1024 links in the road network (including those which are already car free links) were restricted to access by cars. However, the usual bus routes are kept intact and electric cars, taxis can travel to these links. A similar process as that described for Hasselt city is followed. Figure 5 provides the outputs where traffic volume for the morning peak hour can be compared to see the changes in terms of traffic volume for the morning peak hour.

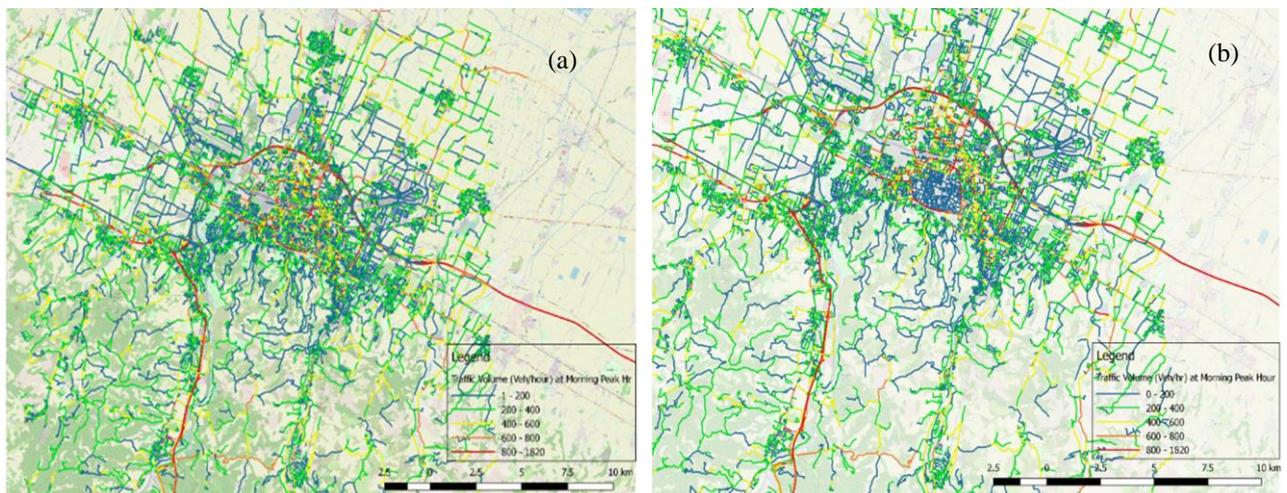

Figure 5: Morning peak hour traffic volume (vehicles/hr) on reduced Bologna Network- (a)Base case (b) Car Access Restricted Policy

The comparison of figure 5a and 5b reveals contrasting results compared to the case of Hasselt. In the case of Bologna, all the roads where car traffic is restricted have very low traffic volume (first category), even though the traffic volume on the ring road has significantly increased. This is because, while in the case of Hasselt the inner ring road itself is restricted for car access, and therefore not connected with other major roads, in the case of Bologna, only the links inside the ring roads are restricted for car access, and therefore, the restriction does not cause any significant loss of connectivity of the road network. Individuals visiting locations inside the ring road can still travel on the ring road to park their vehicle at a decent location. Moreover, car traveller who can previously use links inside ring road to cross the ring are now detouring by moving on the ring road to reach their destination. Furthermore, there is no notable difference found on motorways and other primary arterial roads. In addition to this, we observed the following major changes:

a) Several drivers have detoured their routes as they are now using longer routes in comparison with the base case in which car drivers were using more direct routes. It is noted that the travel time of the car drivers has increased by 13% on average
b) Due to the introduction of a very high penalty for using the road inside the ring, 10% of the car trips toward the inner city have reduced. This 10% of car trips have been distributed as 8% to PT, 1% to bicycle, and 1% to walking. Furthermore, around 2% of individuals have changed their location of activities from the inner-city area to other areas in Bologna
c) The majority of the individuals that have changed their travel mode because of this policy belong to the student population and low income categories.

Figure 6 presents the overall mode share differences between the base case and this policy. The largest difference in percentage points (-1.3%) is noted for car passengers, followed by car drivers and other mode users, and it is translated in terms of higher use of buses followed by bicycles. In relation to absolute numbers, PT trips are increased from 0.337 million trips/day to 0.360 million trips/day trips because of this policy (i.e. a difference of 23000 trips approximately in a day). Car_driver and car_passenger trips are decreased from 0.382 million trips/day to 0.358 million trips/day (i.e. a decrease of 25000 trips/day). This is a significant number of decreases in car trips/day due to the strict nature of this policy.

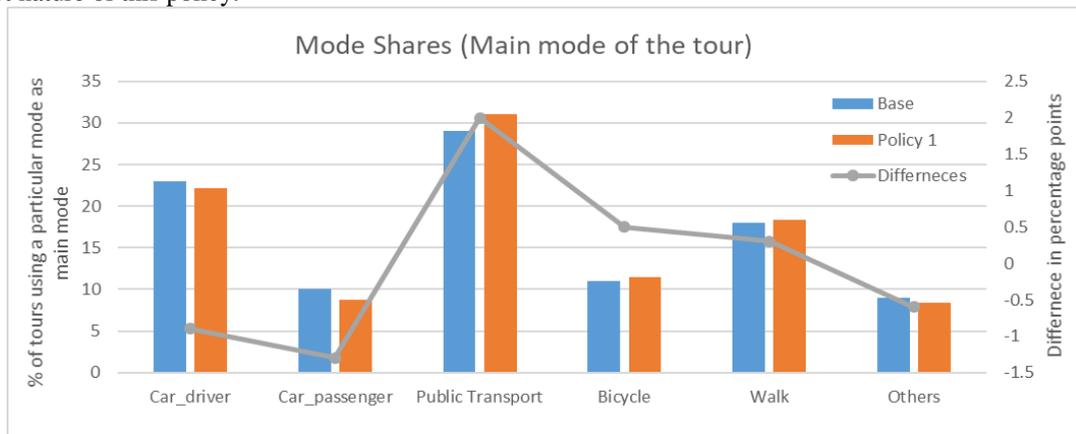

*Figure 6: Mode shares as main travel mode of the tour (along with differences with base case)*

### 4.1.2 Enhancement of Bus Services

The second policy scenario tested for Bologna is based on an enhancement of bus services. In the current situation, the average amount of time people wait at a stop or station for public transit is 12 min, while 16% of riders wait for over 20 minutes on average every day. The average distance people usually ride in a single trip with public transit is 5.4 km, while 7% travel for over 12 km in a single direction [*Moovit, 2019*]. Keeping this context, it is considered that all buses in the fleet are converted into electric and bus routes that are serving the inner area/ring of the city now have their frequency in a manner that waiting times are reduced by 50% especially in the peak times and for off-peak times, waiting times are reduced by 25%. The most crucial component to run this policy is the mode choice model within ABM and MATSIM. Within the mode choice model, the PT utility contains a waiting time variable, whose values are changed during the execution of this policy scenario.

The results indicate that this change increased the PT mode share by only 4% points (i.e. the overall share of PT for this scenario is 33%). Because of this scenario, car use has been reduced but it is not significant as the majority of the bicycle and on-foot travellers have shifted their mode to PT rather than car users. Figure 6 provides a clearer summary of the changes in mode share distribution in this situation. In absolute terms, PT trips increased from 0.337 million trips/day to 0.350 million trips/day (i.e. an increase of 13480 trips in a day), even though this increase is at the cost of a decrease in bicycle trips from 127,827 to 122,017 and walking trips from 209,172 to 197,551. Car traffic is reduced by only 1.5 percentage points, which is around 16,000 trips in magnitude. Similar to Hasselt, the major reason could be that even with low waiting times, the bus is still regarded as a not attractive travel alternative for individuals who are captive car users.

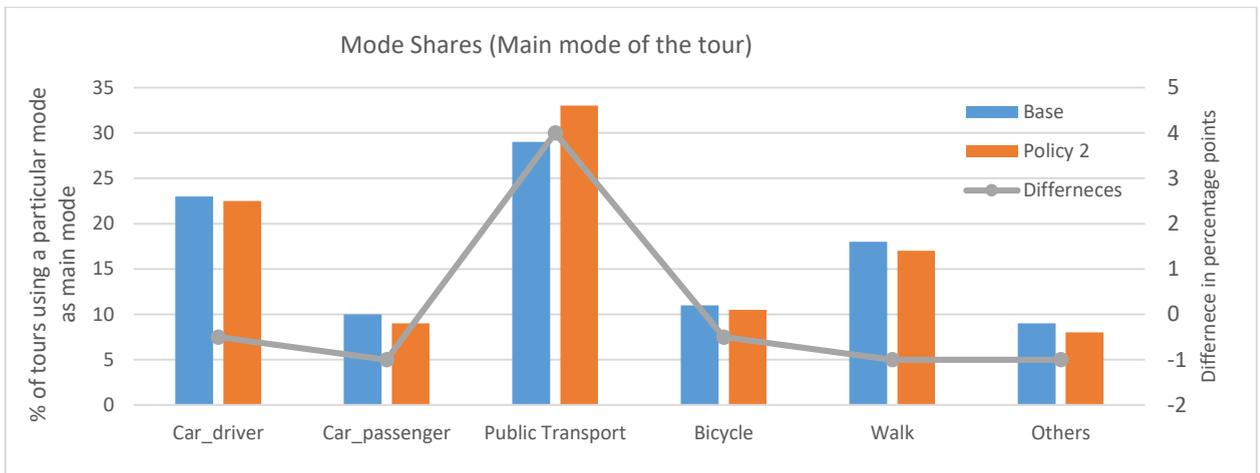

*Figure 7: Mode shares as main travel mode of the tour (along with differences with base case)*

## 5. Discussion

Based on the output generated from the simulations several interesting findings are found. As a general conclusion, car access restriction is found to have more severe impacts on activity-travel behavior with respect to the increase in bus frequency.

Increase in bus frequency as tested for Hasselt and Bologna cases provide negative results as the huge investment by increasing the fleet size will result in attracting a significantly lower number of individuals from car traffic. The ridership of PT increases but as a result of shifting of the individual from other sustainable modes. Similar findings were obtained in earlier studies as well (Adnan and Passani 2017), which further shows that an integrated microsimulation framework is able to produce consistent results.

Car access restriction results in a decrease in modal shares of car drivers and car passengers. In the case of Hasselt, the reduction was higher as compared to the other two cities. This is because in the Hasselt case the access is restricted on the main inner ring which provides connectivity to other roads and major facilities. Conversely, in the case of Bologna the access restriction was limited to links within the city ring road. It should be further noted that, in Hasselt PT network (even in the base case) is not efficient and parameters for waiting time, access time and in-vehicle travel time are lower for PT alternative utility compared to the models estimated for Bologna reflecting the fact that even with some improvements, individuals in Hasselt are more reluctant to shift on PT. Our results indicate that the effect of car restriction as similar to those found in the literature at least effect direction is similar, because, direct comparison cannot be made due to a variety of reasons (i.e. interventions are not exactly the same, significant differences in preferences, etc.)

There are some sources of uncertainty of the simulation framework, such as input data, model parameters, simulation methodology, modelling structure, and the approach followed to obtain the required results. It is not the scope of this study to measure these uncertainties in the results, even though similar studies that investigated uncertainties in similar frameworks mentioned that the order of magnitude of all considered kinds of uncertainty strongly depends on how frequently the alternative is predicted in the choice process. The study of [*Petrik et al., 2018*], about simulation uncertainty (often regarded as simulation error in the literature), estimated modal share percentage uncertainty (in the form of standard deviation) within the range of 0.04% for car traffic. This indicates that the mean values of mode share for car in percent (as reported for two cities in this paper) can be subject to a standard deviation of 0.04%. Apart from this simulation error, the analyses also reported parameter uncertainty. This is done by varying a parameter of the model systematically (i.e. using the concept of multivariate distribution). That analysis obtained a parameter uncertainty for car traffic in the range of 0.4%. In relation to policy scenarios, in which the results are reported for the difference from the base case, we have seen that the percentage point differences are quite lower (i.e.

in the range of 1-2%). It is therefore important that this magnitude of uncertainty is considered when interpreting the effect of policy scenarios.

## 6. Conclusion

The focus of this study is mainly to briefly describe the prototype model (an agent-based integrated microsimulation framework) and the results obtained by performing simulations with a few policy scenarios for two selected cities. Due to significantly detailed requirements of data, which is not always available, a light version of activity-based model is presented that uses simplified data for Hasselt and Bologna. The use of MATSIM in the simulation framework provides flexibility to employ a range of policy scenarios and at the same time, the results are as detailed as possible to obtain the impact of policies on a disaggregate level. The base case scenarios for Hasselt and Bologna are calibrated well enough against available traffic counts, with an error ranging from 0.15 to 0.3. The results of this study indicate that policies considering restrictions in car traffic are more effective with respect to increases in the bus frequency in terms of reducing traffic from the network and also shifting of car drivers/passenger to other modes of travel. This result is largely in agreement with the results reported in other similar studies. Conversely, the enhancement of bus infrastructure in relation to increase the frequency is not able to significantly reduce car traffic. Especially, the improvement of bus infrastructure causes a shifting of bicyclist towards public transport, which is an undesirable result of the policy if the objective is to improve sustainability and environment. It is therefore of paramount important that these traffic management policies are implemented together reach the most appropriate effect.

## Acknowledgements

This project has received funding from the European Union Horizon 2020 research and innovation programme under grant agreement No 689954. This paper reflects the authors views. The European Commission is not liable for any use that may be made of the information contained therein. The authors wish to thank the Municipality of Bologna and the ARPAE Regional Environmental Protection Agency for proving the traffic counts for the city of Bologna.